\documentclass[%
 reprint,
 superscriptaddress,
 amsmath,amssymb,
 aps,
prb,
]{revtex4-1}

\usepackage{graphicx}
\usepackage{dcolumn}
\usepackage{bm}
\usepackage{xcolor}
\usepackage{algorithm}
\usepackage{algpseudocode}
\usepackage{url} 
\usepackage{makecell}
\usepackage{multirow}
\usepackage{booktabs}
\usepackage{graphicx}
\usepackage{subfig}
\usepackage[normalem]{ulem}

\def\br{{\bm r}}

\def\AAb{{\bm A}}

\def\ace{\varphi}
\def\cAA{\tilde{c}}

\def\calF{\mathcal{F}}

\def\calH{\mathcal{H}}

\newcommand {\braHket}[3]{ \langle #1 | #2 | #3 \rangle }

\newcommand {\der}[2]{\frac{d #1}{d #2} }

\begin{document}

\title{Multilayer atomic cluster expansion for semi-local interactions}

\author{Anton Bochkarev}
\email[]{anton.bochkarev@rub.de}
\affiliation{ICAMS, Ruhr-Universit\"at Bochum, Bochum, Germany}
\author{Yury Lysogorskiy} 
\affiliation{ICAMS, Ruhr-Universit\"at Bochum, Bochum, Germany}
\author{Christoph Ortner}
\affiliation{Department of Mathematics, University of British Columbia, Vancouver, BC, Canada V6T 1Z2}
\author{G\'{a}bor Cs\'{a}nyi}
\affiliation{Engineering Laboratory, University of Cambridge, Cambridge, CB2 1PZ UK}
\author{Ralf Drautz} 
\affiliation{ICAMS, Ruhr-Universit\"at Bochum, Bochum, Germany}
\email[]{ralf.drautz@rub.de}

\date{\today}

\begin{abstract}
Traditionally, interatomic potentials assume local bond formation supplemented by long-range electrostatic interactions when necessary. This ignores intermediate range multi-atom interactions that arise from the relaxation of the electronic structure. Here, we present the multilayer atomic cluster expansion (ml-ACE) that includes collective, semi-local multi-atom interactions naturally within its remit. We demonstrate that ml-ACE significantly improves fit accuracy compared to a local expansion on selected examples and provide physical intuition to understand this improvement.
\end{abstract}

\pacs{Valid PACS appear here}
\maketitle

Recent years have seen tremendous progress in modelling atomic interactions~\cite{Behler07,Bartok10,Shapeev16,Willat18,Pun19,Drautz19,Drautz06}. State of the art machine learning interatomic potentials interpolate reference data from high-throughput electronic structure calculations with errors on the order of meV/atom\cite{behler_machine_2021,hart_machine_2021,Musil2021}. Commonly the energy or other atomic quantities are represented as a function of the local atomic environment enclosed within a cutoff radius centered on each atom. Contributions to the energy from electrostatics cannot be partitioned into local atomic environments and methods to incorporate such long-range interactions efficiently have been developed \cite{grisafi_incorporating_2019}, including self-consistent models that mimic the charge transfer of the underlying electronic structure calculations \cite{Ko21,Xie2020}. 

However, the true potential energy from electronic structure calculations contains contributions that evade a local chemical description and cannot be captured by long-range electrostatic models either, even if self-consistent charge transfer is included. We introduce the term ``semi-local'' for interactions that reach significantly beyond the local atomic environment but are not directly associated to long-range charge transfer or directed bond formation. Semi-local interactions are ubiquitous in density functional theory (DFT) and arise from the relaxation of the electronic structure, yet they have not been discussed in the context of machine learning potentials. Examples are the change of interaction in small clusters with size that approach bulk interactions only slowly; intra-atomic occupation changes upon hybridization, such as the promotion of electrons in carbon from $s$ to $p$ states, that alter the carbon bonding characteristics; variation in atomic hybridization in different atomic environments that induces metal-insulator transitions with consequences for the decay of the density matrix and related bond formation; electronic states along one-dimensional chains that can extend far beyond the local chemical environment.

We build our analysis of semi-local interactions on a local description of the electronic structure and expand the DFT energy with respect to modifications of the density matrix \cite{foulkes_tight-binding_1989,finnis_interatomic_2003-1,Drautz11,drautz_bond-order_2015}, 
\begin{equation}
E = E_0 + \mathrm{tr}\left(\pmb{H}\Delta \pmb{\rho}\right) + \mathrm{tr}\left(\pmb{J}\Delta \pmb{\rho} \Delta\pmb{\rho}\right) + \dots \,,
\end{equation}
with $E_0 = E[\pmb{\rho}_0]$, the Hamiltonian matrix $\pmb{H}$ and the density matrix $\pmb{\rho} = \pmb{\rho}_0 + \Delta \pmb{\rho}$. For making contact with interatomic interactions we assume orbitals $\alpha,\beta,\ldots$ that are localized on atoms $i,j,\dots$ and the density matrix elements are given as $\rho_{i \alpha j \beta} = \braHket{i \alpha}{\hat{\rho}}{j \beta}$. 
The spectrally resolved density matrix $n_{i \alpha j \beta}(E_F) = \der{\rho_{i \alpha j \beta}}{E}(E_F)$
is linked to the Hamiltonian through the generalized moments theorem \cite{cyrot-lackmann_electronic_1967,Drautz06}
\begin{align}
    &\int E^N n_{i \alpha j \beta}(E)\, dE =  \braHket{i \alpha}{\hat{H}^N}{j \beta} \nonumber \\ 
    =&
    \sum_{k \gamma l \delta} H_{i \alpha k \gamma} H_{k \gamma l \delta} H_{l \delta \dots} \dots H_{\dots j \beta} \,, \label{eq:mom}
\end{align}
with $H_{i \alpha j \beta} = \braHket{i \alpha}{\hat{H}}{j \beta}$ and where we further took the orbitals to be orthonormal and complete. This enables to reconstruct the density matrix as a linear combination of products of 
Hamiltonian matrix elements of varying order\cite{pettifor_new_1989,Drautz06,thomas2021rigorous},
\begin{equation}
    \rho_{i \alpha j \beta} = \chi_1 {H}_{i \alpha j \beta} + \chi_2 \sum_{k \gamma} {H}_{i \alpha k \gamma} {H}_{k \gamma j \beta} + \dots  \,,
\end{equation}
where the response functions $\chi_N = \chi_N(E_F)$ depend on the Fermi level. 
In tight binding approximation off-diagonal Hamiltonian matrix elements $H_{i \alpha j \beta}$ with $i \neq j$ depend only weakly on electronic redistribution, while the diagonal elements  $H_{i \alpha i \alpha}$ follow the effective one-particle potential and adjust to optimize energy for hybridization and charge transfer \cite{finnis_interatomic_2003-1}.  The detailed change  $\Delta H_{i \alpha i \alpha}$ is a function of the local atomic environment of atom $i$, which may be understood from the moments theorem Eq.(\ref{eq:mom}) applied to the local density of states $ n_{i \alpha i \alpha}(E) $. For example, in a metal often charge transfer is negligible and $\Delta H_{i \alpha i \alpha}$ adjusts to variations in the local density of states to keep the number of electrons on an atom constant. 

\begin{figure*}[ht!]
    \includegraphics[width=2.0\columnwidth]{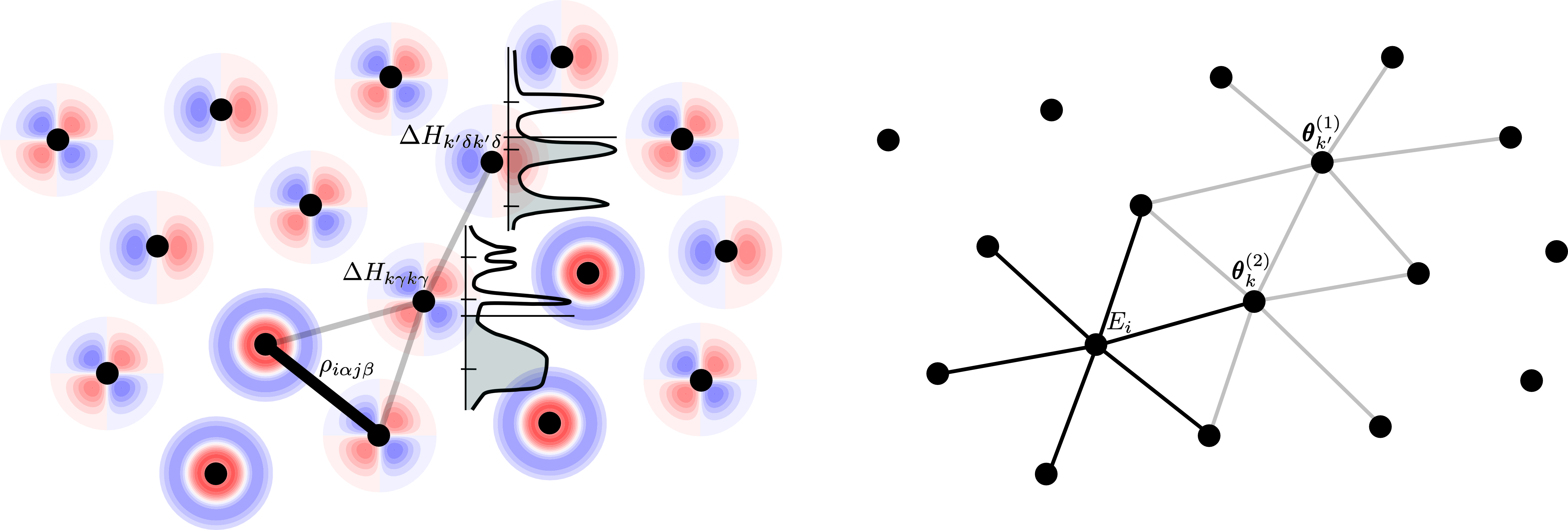}
    \caption{Illustration of semi-local interactions in electronic structure calculations. The density matrix element $\rho_{i \alpha j \beta}$ between atoms $i$ and $j$ is modified by changes in the onsite levels $\Delta H_{k \gamma k \gamma}$ of neighboring atoms $k$. The changes in the onsite levels depend on the Fermi level and the local density of states $n_{k \gamma k \gamma}$, which in turn is a function of the local environment of atom $k$ and depends on the onsite levels of further distant atoms. The onsite levels and density of states on further distant atoms are then again a function of their local environment (left). Abstraction of the electronic interactions in ml-ACE. The energy $E_i$ of atom $i$ depends on the state of the neighboring atoms through the indicator field $\pmb{\theta}^{(2)}_k$ on atom $k$. The state of each neighboring atom depends on the local atomic environment of the atom, where the state of the neighboring atom depends on a further indicator field $\pmb{\theta}^{(1)}_{k'}$ (right).
    }
    \label{fig:rACE}
\end{figure*}

\begin{figure}[hb!]
    \includegraphics[width=\columnwidth]{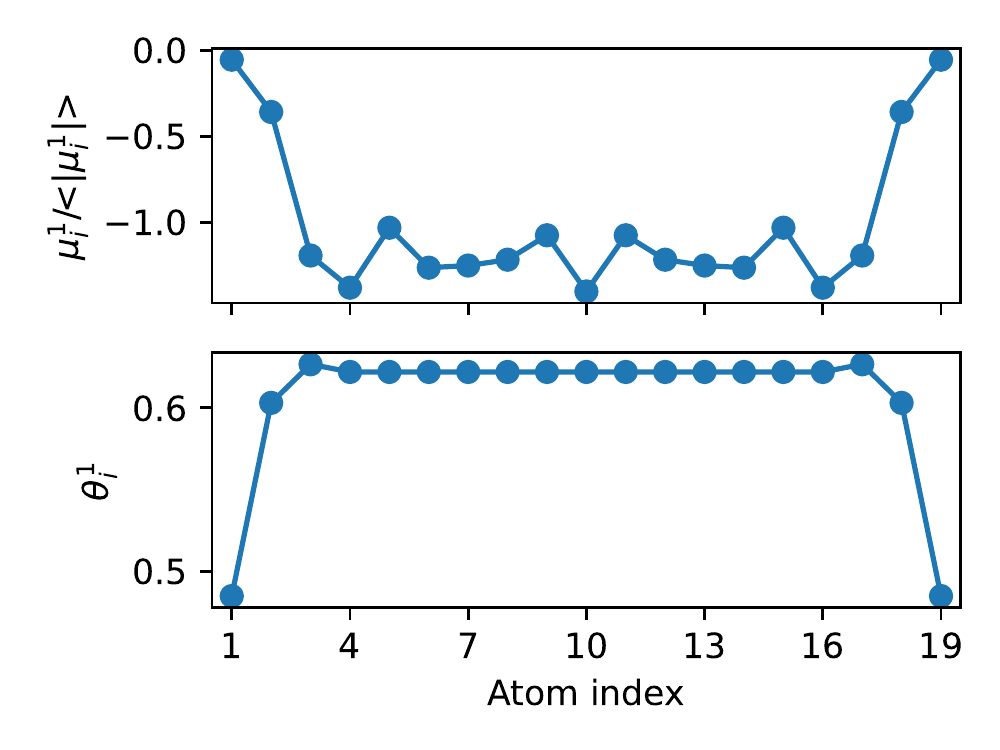}
    \caption{Upper panel shows normalized first moment of the atomic density of state along 19-atom linear Cu chain. Lower panel shows values of the indicator for each atom from the first layer of a two layer ACE model. 
    }
    \label{fig:linear}
\end{figure}

Therefore electronic relaxation $\Delta H_{k \gamma k \gamma}$ on atom $k$ will affect the density matrix between atoms $i$ and $j$ to lowest order as
\begin{equation}
\Delta \rho_{i \alpha j \beta} \propto H_{i \alpha k \gamma} \Delta H_{k \gamma k \gamma} H_{k \gamma j \beta} \,. \label{eq:dE}
\end{equation}
In general the off-diagonal Hamiltonian elements decay rapidly with distance between atoms $i$ and $j$ and local neighbors $k$ have the strongest effect on the strength of bond $i-j$, cf. Refs.\citenum{finnis_interatomic_2003-1,Drautz06,thomas2021rigorous}. The onsite Hamiltonian matrix elements on atom $k$, in turn, are determined to minimize the energy, which is a function of the density of states $n_{k \gamma k \gamma}(E)$ that depends on the environment of atom $k$, which includes atoms $k'$ further distant from the bond $i-j$, etc. Therefore the bond $i-j$ is modified by a decaying cascade of modifications on neighbors of neighbors, resulting in semi-local interactions as illustrated in Fig.~\ref{fig:rACE}. We show the first moment of the atomic density of states $\mu_i^{(1)} = \sum_{\alpha} H_{i \alpha i \alpha}$ of a linear chain of Cu atoms in Fig.~\ref{fig:linear}. 

Here, we present a general framework for integrating semi-local interactions from electronic structure calculations efficiently into machine-learned potentials (MLPs). This is achieved by first describing the atoms according to their local atomic environment before modelling the chemical bond formation between the atoms. In a first step, atoms condense information about their local environment into an indicator field. Next, atoms condense indicator fields of their neighbors into their own indicator field, see Fig.~\ref{fig:rACE}. This is repeated and with each layer information from atoms at larger distances is incorporated, mimicking the electronic structure relaxation cascade in self-consistent calculations and enabling the description of collective interactions that extend multiple times beyond the local cutoff radius.

The Atomic Cluster Expansion (ACE)\cite{Drautz19,Dusson20,lysogorskiy_performant_2021,bochkarev_efficient_2022} provides a complete and efficient local representation of the energy as a sum over atomic contributions,
\begin{equation}
E_i = \calF(\ace_i^{(1)}, \dots,\ace_i^{(P)} ) \,,\label{eq:EF}
\end{equation}
where $\calF$ is a general non-linear function. Each atomic property $\ace^{(p)}_i$ is given by a linear expansion, 
\begin{equation}
    \ace_i^{(p)} =     \sum_{\pmb{v}} \cAA^{(p)}_{\pmb{v}} {\AAb}_{i\pmb{v}}  \label{eq:rho} \,,
\end{equation}
with expansion coefficients $\cAA^{(p)}_{\pmb{v}}$ and multi-atom basis functions ${\AAb}_{i\pmb{v}}$.

Other variables than the atomic positions may be taken into account\cite{Drautz20}. For example, charges or atomic magnetic moments may be assigned to the atoms and the resulting energy depends on the values of these variables. To this end the state  $\pmb{\sigma}$ of the system is introduced that collects all necessary variables and comprises edges and vertices. 
We extend ACE to include semi-local interactions by extending the description of the state of an atom by an indicator field $\pmb{\theta}$. The field is general, it may consist of different contributions and comprise several scalar, vectorial or tensorial elements. For an expansion on atom $i$, the state of a neighboring atom $j$ is defined as
\begin{equation}
\sigma_{ji} = (z_j, \br_{ji},\pmb{\theta}_j ) \,,
\end{equation}
while the state of atom $i$ is given by
\begin{equation}
\sigma_{ii} = ( z_i,\pmb{\theta}_i)  \,,
\end{equation}
where $z_i, \br_i$ denote, respectively, the chemical element and position of atom $i$. For simplicity we left out possible dependencies on charge, magnetism, etc. These variables are absorbed into a complete set of single-particle, respectively single-bond basis functions $\phi_{v}(\sigma_{ji})$, from which the atomic base is computed
\begin{equation}
A_{iv} = \sum_j \phi_{v}(\sigma_{ji}) \,,
\end{equation}
and $A^{(0)}_{iv} = \phi^{(0)}_{v}(\sigma_{ii})$ for contributions that depend only on atom $i$. The products of the atomic base ${\AAb}_{i\pmb{v}} = A^{(0)}_{iv_0} \prod_{n}A_{i v_n}$ of various order form a complete set of basis functions and enable Eq.(\ref{eq:rho}) to represent any atomic function of $\pmb{\sigma}$. The expansion remains essentially unchanged when carried out for vectorial or tensorial objects \cite{Drautz20}.

Often it is pragmatic to choose product basis functions of the form
\begin{equation}
\phi_v(\sigma_{ji}) = e_{\kappa}(z_j) R_{nl}(r_{ji}) Y_l^m(\hat{\pmb{r}}_{ji}) T_k(\pmb{\theta}_j) \,, \label{eq:basis}
\end{equation}
where $v$ collects the necessary indices and the functions $T_k$ are complete in the space spanned by the indicator fields~\cite{Drautz20}. The indicator field then depends on indicator fields of other atoms itself, etc. such that a recursive dependence is established as illustrated in Fig.~\ref{fig:rACE}. 

For an expansion with several layers, the indicator fields $\pmb{\theta}_j^{(n)}$ and expansion coefficients $\cAA^{(p,n)}$ differ from layer to layer and atomic properties of layer $n$ are represented as $\ace_i^{(p,n)} =   \sum_{\pmb{v}} \cAA^{(p,n)}_{\pmb{v}} {\AAb}_{i\pmb{v}}(\pmb{\theta}^{(n)})$.
The indicator field in the next layer $n+1$ is then obtained from a non-linear function as Eq.(\ref{eq:EF}),
\begin{equation}
\pmb{\theta}_i^{(n+1)} = \calH(\ace_i^{(1,n)}, \dots,\ace_i^{(P,n)} ) \,.\label{eq:theta}
\end{equation}
Layer $n_{max}$ is the output layer, while the input layer $n = 0$ is initialized via $T_k(\pmb{\theta}^{(0)}) = 1$. The indicator fields in general carry particular symmetries, most importantly covariance under rotation and inversion, which requires symmetrization.~\cite{Drautz20,Dusson20} We denote the resulting model the {\em multilayer Atomic Cluster Expansion} (ml-ACE).

By adding new features in the form of indicator fields ml-ACE becomes high-dimensional and sparse basis sets are required for converging ml-ACE. This may be achieved, e.g, with sparsification algorithms, but in the present work we are guided by physical intuition and hierarchical analysis. The further a layer is away from the final layer, the smaller its impact will be on the final prediction, i.e. the details are often lost in the distance. This means that the complexity of the indicator field needs to be varied across the layers for best performance and efficient convergence.
Because of the layered structure of ml-ACE, force gradients may be obtained efficiently and the computational expense is linear or less with the number of levels, as some basis functions can be reused, in particular the spherical harmonics.

The ml-ACE may be adapted to represent various message passing networks architectures. In fact, general message passing networks may be obtained as special cases of ml-ACE, see~\citet{GaborNetworks} and \citet{nigam2022unified}. For example, \citet{Thomas18} and the closely related NequIP network\cite{Batzner21} may be cast in the form of a particular low order ACE on each layer.
 
In the following we present results for a basic version of ml-ACE with scalar (invariant) indicator functions. In simple tight-binding models the difference of the onsite levels are fixed and their shifts may be characterized by the first moment of the atomic density of states $\mu^{(1)}_i = \sum_{\alpha} H_{i \alpha i \alpha}$ using Eq.(\ref{eq:mom}). We therefore take a single indicator variable, which in addition we assume to be rotationally invariant, and that is represented as a function of the local atomic environment by ACE as Eq.(\ref{eq:rho}). We expect a linear change of the energy with indicator field for small changes, Eq.(\ref{eq:dE}). To make contact with traditional neural networks, we further choose $\calH = \tanh$. The indicator function is then transformed as $\theta_i = \tanh ( \ace_i )$. 
The order parameters becomes part of ACE by a suitable choice of single bond basis functions and for a single order parameter we choose the basis functions $T_k$ as Chebyshev polynomials of the first kind.

We demonstrate the performance of the ml-ACE with examples of two distinct cases, namely small metallic Cu clusters and 10 small organic molecules from the revMD17 dataset.~\cite{revMD17}

\begin{figure}[h!]
    \includegraphics[width=\columnwidth]{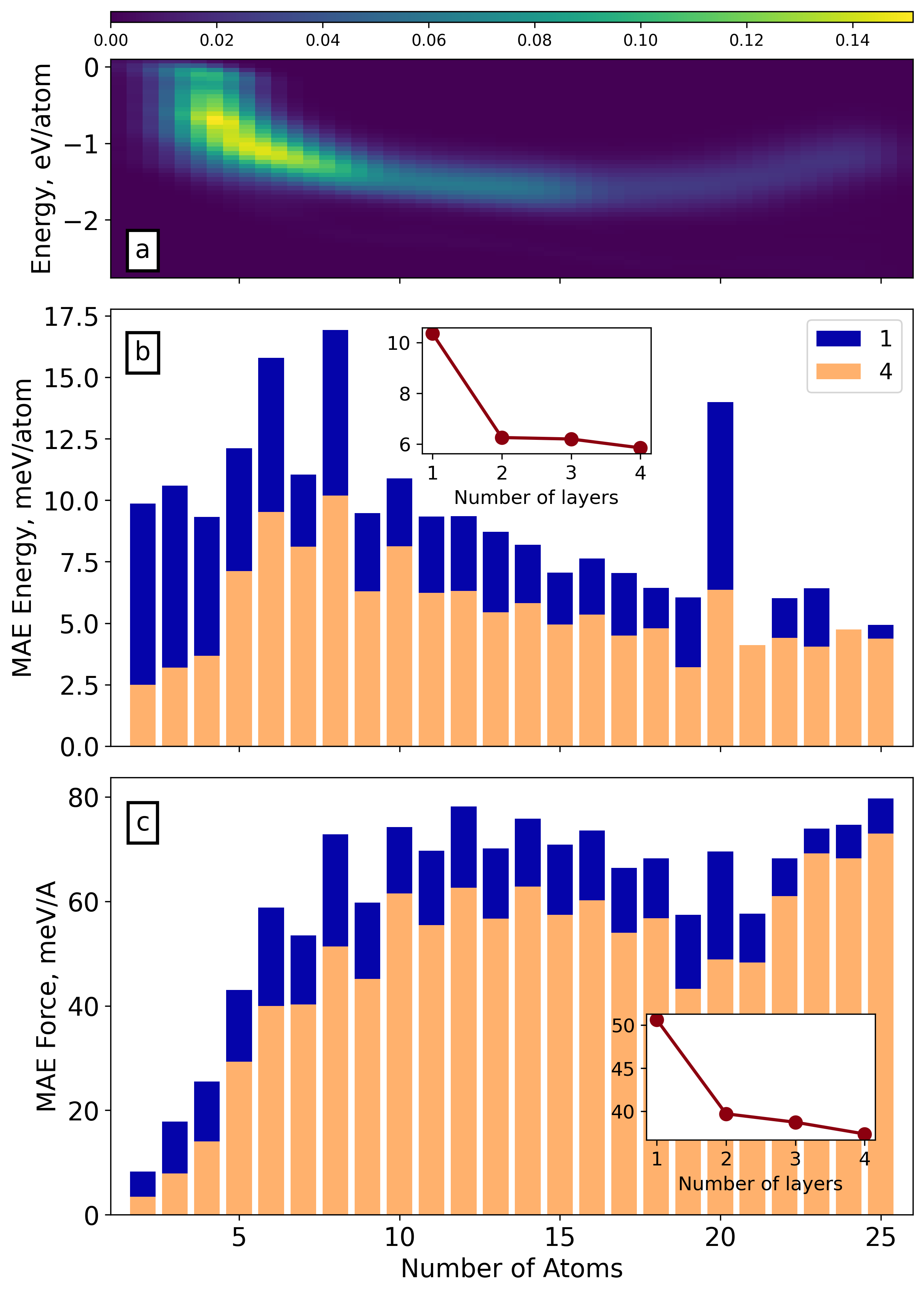}
    \caption{{(a)} Distribution of structures in the Cu clusters test set as function of energy and number of atoms. Color shows the fraction of total 62978 structures with given energy and size. (b),(c) Test MAE for predicting energies and forces of small Cu clusters using ml-ACE with various number of layers, where 1 denotes a standard non-layer ACE. Bars show predictions as a function of cluster size. Insets demonstrate convergence of the ml-ACE predictions averaged over the entire test set.}.
    \label{fig:cluster}
\end{figure}

In a small cluster of $N$ atoms to which one more atom is added, the extra atom can significantly change the interaction between all atoms in the cluster, implying that $(N+1)$-body interactions are necessary, and $(N+2)$-body interactions when a further atom is added. One expects a slow convergence to bulk interactions only as $N^{-1/3}$, simply as in a compact cluster the number of surface atoms (with a significantly modified density of states) is proportional to $N^{2/3}$.
We employ a dataset of nearly 70000 small Cu clusters. The energies and forces of the cluster configurations were computed with FHI-aims \cite{FHIaims,FHIaims2} using a tight basis set. The clusters contain from 2 to 25 atoms. Most of reference clusters were generated by randomly placing atoms inside a sphere of a given radius and ensuring that the distance between any pair of atoms is not smaller than 80\% of the bulk nearest neighbor distance. In addition cuts from various crystalline bulk phases were used as well as cluster geometries obtained with empirical potentials. For clusters of up to four atoms the positions were varied systematically to sample the complete configuration space. The distribution of the size and energies of the clusters is shown in Fig.~\ref{fig:cluster}a. All the cluster structures were distinctly different, none of the clusters were relaxed and in particular we did not use data along molecular dynamics trajectories to avoid bias because of the long de-correlation times. We randomly selected 10~\% of the dataset for training and the remaining 90~\% of the clusters for testing.

Fig.~\ref{fig:cluster} shows the convergence of the mean average error (MAE) of forces and energies of ml-ACE for a single, scalar invariant indicator variable. A major improvement of nearly a factor of two is achieved compared with the standard single layer ACE model. Further increase in the model depth yield smaller yet consistent improvements, in accordance with expected rapid convergence with the number of layers \cite{thomas2021rigorous}. Small clusters with up to four atoms and larger clusters with about 15 or more atoms show errors of only a few meV. The description of small clusters with up to four atoms benefits from the most accurate description of the many-body potentials up to body-order four enabled by ml-ACE. Larger clusters from about 15 atoms benefit from the accurate description of semi-local collective interactions. The largest errors are observed for cluster sizes between about 5 to 10 atoms.

We illustrate the correlation between indicator field and electronic structure in 
Fig.~\ref{fig:linear}, where we show the first moment of the atomic density of states along a 19-atom linear chain of Cu atoms and compare this to the values of the scalar indicator field of a two layered ACE model. The indicator field picks up the largest deviations at the boundary of the chain. We note that atomic linear chains were not part of the training set. 

Extended interactions also contribute to bond formation in small molecules. Table~\ref{table:ethanol} shows the convergence of ml-ACE models for an ethanol molecule from the revised MD17 dataset.~\cite{revMD17, Christensen_2020} Similar to small clusters, adding a second layer gives the biggest performance gain, while every further layer results in smaller yet consistent improvement. Thus, for all other molecules we trained 4-layer ACE models and the corresponding performance metrics are summarised in Table~\ref{table:md17} (model details are given in Supplementary Materials). The resulting models outperform most of the machine learning potentials~\cite{Christensen_2020,Christensen_2020_1,Kovacs2021} with the exception of the equivariant neural network potentials NequIP~\cite{Batzner21}. 
Note that we only compare to models trained on the revised MD17 dataset and Table~\ref{table:md17} shows comparison to the closely related linear ACE model and the best performing model NequIP. For other related models see Refs.~\onlinecite{Kovacs2021,Anderson19,Schutt2021,NewtonNet21}.
Performance improvement of the ml-ACE via semi-local information is evidenced by comparing to the linear ACE models. These models contain an order of magnitude more independent basis functions yet they are outperformed by ml-ACE for all molecules. On the other hand, NequIP models improve over ml-ACE for almost all molecules with varying performance differences.
The largest absolute difference in force MAE ($\sim$ 6.5 meV/$\text{\AA}$) is observed for the aspirin molecule, while the smallest difference is observed for benzene (0.1 meV/$\text{\AA}$) with the ml-ACE model being slightly more accurate. This discrepancy can be understood from the structure of the molecules.  Charge density is nearly uniform in the benzene molecule and therefore small variations in the atomic environments during dynamics do not introduce significant charge redistribution and the semi-local information provided by a scalar invariant indicator is sufficient.
On the contrary, in the aspirin molecule charge distribution is highly non-uniform and changes significantly during dynamics~\cite{Hauf2019}. This leads to a variation of $s$-$p$ hybridization on the atoms and non-uniform onsite level modifications for $s$ and $p$ orbitals. The NequIP potential benefits from propagating additional non-scalar equivariant information across the layers which scalar invariant indicators of the current implementation of ml-ACE are unable to capture (see also Supplementary Materials for more details).

\begin{table}
\centering
\caption{Performance of the ACE models with various number of layers for the ethanol molecule from revised MD17 dataset.~\cite{revMD17}}
\label{table:ethanol}
\begin{tabular}{ccc}
\hline
\hline
\# of layers & MAE E, meV & ~MAE F, meV/$\text{\AA}$~  \\
\hline
1            & 1.9        & 9.8                \\
2            & 1.0        & 6.2                \\
3            & 0.8        & 5.4                \\
4            & 0.7        & 4.4               \\
\hline
\hline
\end{tabular}
\end{table}

\begin{table}[]
\caption{MAE for energies (E, meV) and forces (F, meV/\AA) evaluated on the test set for the 4-layer ACE models and compared to the linear ACE models~\cite{Kovacs2021} and NequIP~\cite{Batzner21} (we show values corresponding to $l=3$).}
\label{table:md17}
\begin{tabular}{ccccc}
\hline
\hline
 Molecule                       &   & ml-ACE (this work) & Linear ACE & NequIP \\ \hline
\multirow{2}{*}{Aspirin}        & E & 4.7      & 6.1      & 2.3\\
                                & F & 14.9     & 17.9     & 8.5 \\ \hline
\multirow{2}{*}{Azobenzene}     & E & 2.3      & 3.6      & 0.7 \\
                                & F & 7.7      & 10.9     & 3.6 \\ \hline
\multirow{2}{*}{Benzene}        & E & 0.02     & 0.04     & 0.04\\
                                & F & 0.2      & 0.5     & 0.3 \\ \hline
\multirow{2}{*}{Ethanol}        & E & 0.7     & 1.2      & 0.4 \\
                                & F & 4.4     & 7.3      & 3.4 \\ \hline
\multirow{2}{*}{Malonaldehyde}  & E & 1.3      & 1.7      & 0.8 \\
                                & F & 8.6      & 11.1     & 5.2 \\ \hline
\multirow{2}{*}{Naphthalene}    & E & 0.8      & 0.9      & 0.2 \\
                                & F & 3.9      & 5.1      & 1.2 \\ \hline
\multirow{2}{*}{Paracetamol}    & E & 3.2      & 4.0    & 1.4   \\
                                & F & 10.7     & 12.7  & 6.9  \\ \hline
\multirow{2}{*}{Salicylic acid} & E & 1.5      & 1.8      & 0.7 \\
                                & F & 7.7      & 9.3      & 4.0 \\ \hline
\multirow{2}{*}{Toluene}        & E & 0.8      & 1.1      & 0.3 \\
                                & F & 4.3      & 6.5      & 1.6 \\ \hline
\multirow{2}{*}{Uracil}         & E & 0.6        & 1.1    &  0.4\\
                                & F & 4.0        & 6.6    & 3.2 \\
\hline
\hline
\end{tabular}
\end{table}

To conclude, semi-local interactions are an integral contribution in electronic structure calculations that are used for training machine learning potentials. Here we introduce ml-ACE that efficiently captures electronic structure relaxation in self-consistent schemes such as DFT. We show that the indicator fields from ml-ACE can be understood from physical and chemical intuition and note that message passing networks may be cast in the form of ml-ACE. 

We demonstrate ml-ACE numerically for the special case of a single scalar invariant per atom. This allows us to reduce errors for small metallic clusters and molecules to about 50\% as compared to single layer models. Employing only a single scalar invariant is also the main limitation of the numerical implementation presented here. Including several indicator variables with non-scalar rotational characteristics, corresponding to $p$, $d$, etc. onsite Hamiltonian elements are the necessary next steps for reducing the remaining errors further. As the focus of our work is on semi-local interactions, our implementation further neglected long-range interactions due to charge transfer. Both contributions, equivariant, angularly dependent indicator fields associated to $p$ and $d$-valent onsite levels as well as charge transfer need to be taken into account for our next implementation of ml-ACE. 

\section*{Acknowledgement}

R.D. acknowledges funding through the German Science Foundation (DFG), Projects No. 405621217 and No. 403582885. Y.L. acknowledges funding through the German Science Foundation (DFG), Project No. 405602047. R.D. acknowledges computational resources of the research center ZGH, Ruhr-University Bochum, Germany.

\end{document}



\title{Supplemental material: Multilayer atomic cluster expansion for semi-local interactions}

\author{Anton Bochkarev}
\email[]{anton.bochkarev@rub.de}
\affiliation{ICAMS, Ruhr-Universit\"at Bochum, Bochum, Germany}
\author{Yury Lysogorskiy} 
\affiliation{ICAMS, Ruhr-Universit\"at Bochum, Bochum, Germany}
\author{Christoph Ortner}
\affiliation{Department of Mathematics, University of British Columbia, Vancouver, BC, Canada V6T 1Z2}
\author{G\'{a}bor Cs\'{a}nyi}
\affiliation{Engineering Laboratory, University of Cambridge, Cambridge, CB2 1PZ UK}
\author{Ralf Drautz} 
\affiliation{ICAMS, Ruhr-Universit\"at Bochum, Bochum, Germany}
\email[]{ralf.drautz@rub.de}

\date{\today}
             
\maketitle

\section{Model and fitting details}
Multilayer ACE models are trained via minimizing a loss function of the following form
\begin{eqnarray}
\label{eq:loss_function}
    \mathcal{L} &=& (1-\kappa)\sum_{n=1}^{N_\mathrm{struct}} w_n^{(E)} \left(\frac{E_n^\mathrm{ACE}-E_n^\mathrm{ref}}{n_\mathrm{at,n}}\right)^2  \nonumber \\ 
    &+& \kappa \sum_{n=1}^{N_\mathrm{struct}} \sum_{i=1}^{n_\mathrm{at,n}} w_{ni}^{(F)} \left(\bm{F}_{ni}^\mathrm{ACE}-\bm{F}_{ni}^\mathrm{ref}\right)^2, 
\end{eqnarray}
where  $\kappa$ is a trade-off between energy and force contribution, $N_\mathrm{struct}$ is the number of structures employed in the parameterization, $n_\mathrm{at,n}$ the number of atoms in structure $n$, and $w_n^{(E)}$ and $ w_n^{(F)}$ are per-structure and per-atom weights for the energy and force residuals, which were set to 1 for every structure and normalized by the number of structures and atoms respectively. 
For small molecules, we select structures for fitting according to the first split from the revMD17 dataset.~\cite{revMD17} Multilayer ACE models were implemented and fitted within tensorpotential package~\cite{bochkarev_efficient_2022} and Table~\ref{table:modelparam} summarizes the model hyper-parameters used for each system.

\begin{table*}
\centering
\caption{Details of the potential configurations used in this work.}\label{table:modelparam}
\begin{tabular}{lcccc}
System                       & \# of layers    & cutoff,~$\AA$      & $\kappa$ & \# functions/element   \\ 
\hline
\hline
\multirow{4}{*}{Cu clusters} & 1                  & \multirow{4}{*}{8} & \multirow{4}{*}{0.1} & 993                    \\
                             & 2                  &                    &                    & 1421                   \\
                             & 3                  &                    &                    & 1911                   \\
                             & 4                  &                    &                    & 2401                   \\ 
\hline
\multirow{4}{*}{Ethanol}     & 1                  & \multirow{4}{*}{4} & \multirow{4}{*}{0.99} & 1207                   \\
                             & 2                  &                    &                        & 1363                   \\
                             & 3                  &                    &                        & 2269                   \\
                             & 4                  &                    &                        & 2982                   \\ 
\hline
Aspirin                      & \multirow{2}{*}{4} & \multirow{2}{*}{5} & \multirow{2}{*}{0.99} & 2982  \\
Naphthalene                  &                    &                    &                       & 2922                      \\ 
\hline
Azobenzene                   & \multirow{7}{*}{4} & \multirow{7}{*}{4} & \multirow{7}{*}{0.99} & 2982                       \\
Malonaldehyde                &                    &                    &                       & 2982                     \\
Salicylic acid               &                    &                    &                       & 2982                     \\ 
Toluene                      &                    &                    &                       & 2922  \\
Benzene                      &                    &                    &                       & 2922                      \\
Paracetamol                  &                    &                    &                       & 2906  \\
Uracil                       &                    &                    &                       & 2906                     \\
\hline
\hline
\end{tabular}
\end{table*}

\section{Small molecules}

\begin{figure}
    \includegraphics[width=\columnwidth]{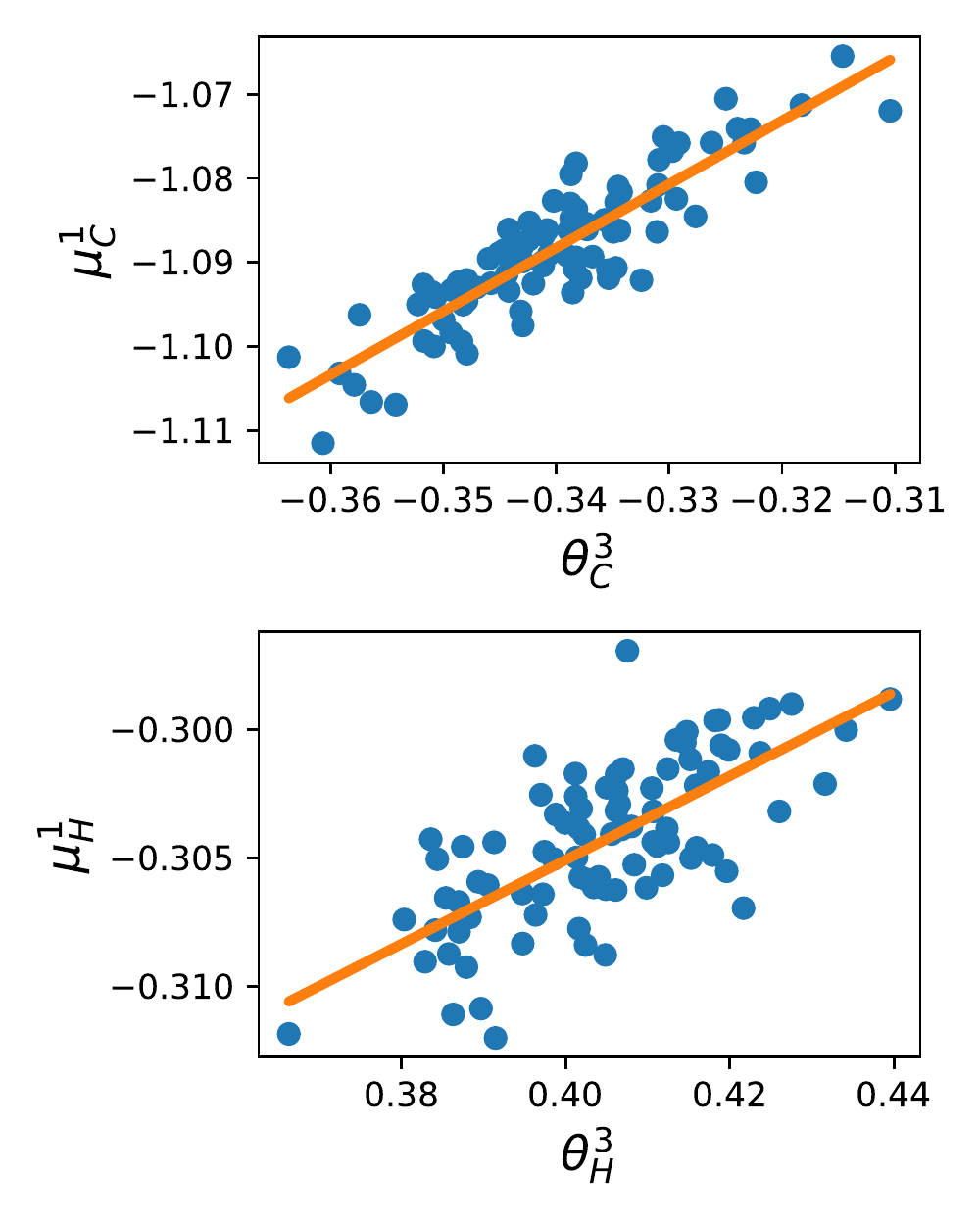}
    \caption{Correlation between first moment $\mu^{(1)}_i$ and indicator value $\theta^{(3)}_i$ for carbon (top) and hydrogen (bottom) atoms in the benzene molecule. Values are computed for 15 randomly selected molecules from the training set.
    }
    \label{fig:benz}
\end{figure}

\begin{figure}
    \includegraphics[width=\columnwidth]{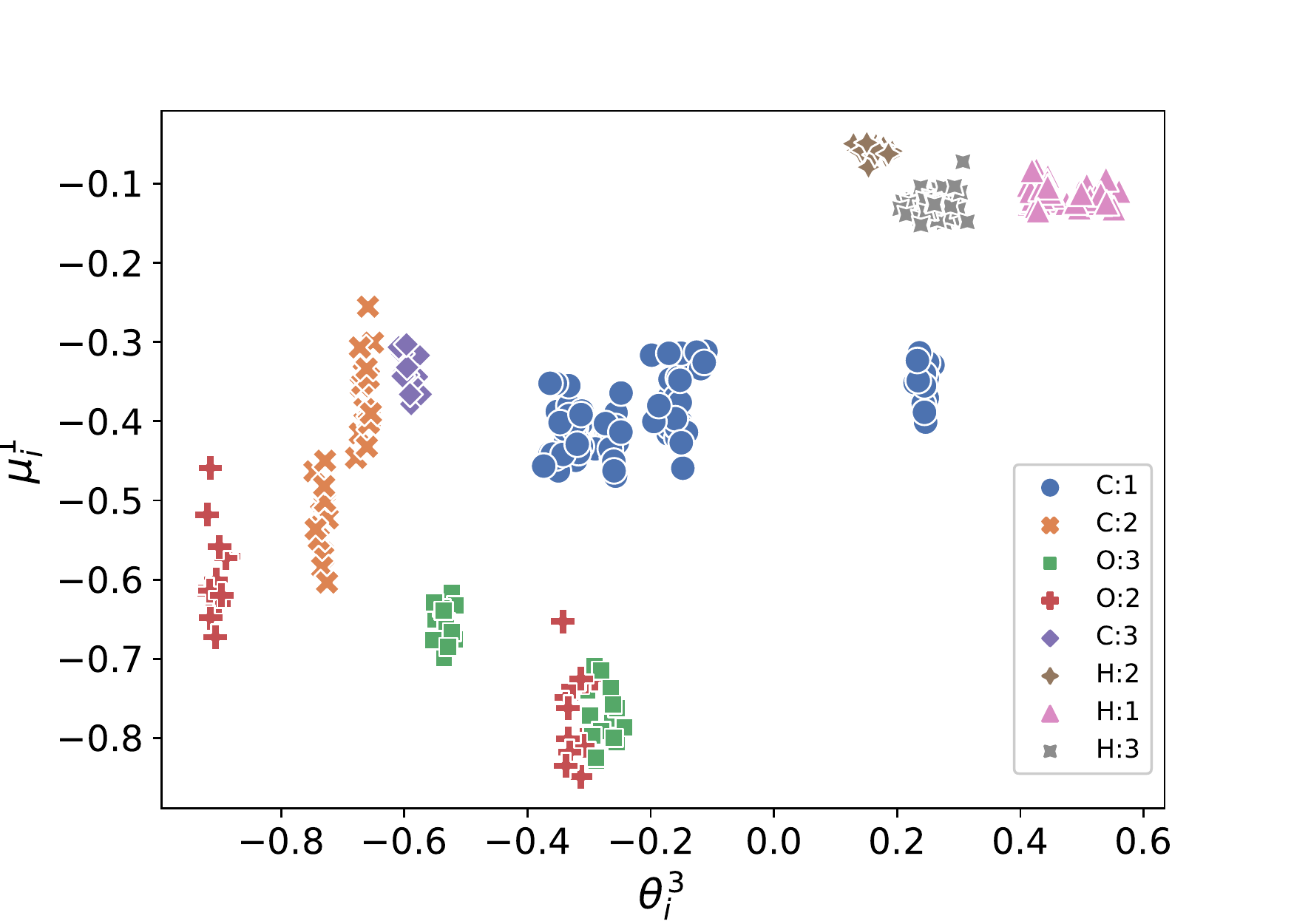}
    \caption{Correlation plot between first moment $\mu^{(1)}_i$ and indicator value $\theta^{(3)}_i$ for atoms in the aspirin molecule. Values are computed for 15 randomly selected molecules from the training set. Index in the legend denotes the group of an atom, where 1 - benzene ring, 2 - acetyl group, 3 - carboxyl group.}
    \label{fig:asprn}
\end{figure}

For understanding the difference in performance of the multilayer ACE models for different molecules we consider two cases - the best performing model for the benzene molecule and the worst performing model for the aspirin molecule. Fig~\ref{fig:benz} shows the correlation between values of the scalar indicator $\theta^{(3)}_i$ of the 4-layer ACE model and the first moment $\mu^{(1)}_i$ of the atomic DOS for each atom type computed for 15 randomly selected molecules from the training set using FHI-aims \cite{FHIaims,FHIaims2} with tight settings. As expected, the electronic distribution in benzene is rather uniform and does not change significantly during dynamics, which is illustrated by the narrow window of values of the first moment. Thus, the invariant scalar indicator is able to capture these small differences via propagating the semi-local information through the local atomic environments, which is illustrated by the linear correlation between values of $\mu^{(1)}_i$ and $\theta^{(3)}_i$. Fig.~\ref{fig:asprn} shows the correlation for 15 randomly selected aspirin molecules, recomputed with FHI-aims with tight settings. Data are grouped according to their type and position in a particular functional group. However, unlike Fig.~\ref{fig:benz}, no clear correlation between value of $\theta^{(3)}_i$ and $\mu^{(1)}_i$ is observed. This implies that in this case the scalar indicator provides incomplete information leading to inferior model performance. 

